\documentstyle[12pt,epsfig]{article}

\newcommand{\myskip}{\vspace{\baselineskip}}
\newcommand{\mysection}[1]{\par\myskip\noindent\textbf{#1}\myskip\par}

\begin{document}

\begin{center}
{\large\textbf{Photoassisted dynamical transport in multiple quantum wells}} 

\myskip 
Rosa L\'opez, David S\'anchez, and Gloria Platero\\
Instituto de Ciencia de Materiales de Madrid (CSIC), 
Cantoblanco, 28049 Madrid, Spain.

\mysection{Abstract}

\parbox{4.5in}{We study the dynamical transport in weakly coupled superlattices
in the presence of intense radiation in the THz regime.
We derive a general model for the time dependent tunneling current within
the Keldysh nonequilibrium-Green-function formalism. 
For the particular situation in which fast scattering procceses
drive the system to local equilibrium in the wells
drastic changes are found in the current vs. voltage curves.}
\end{center}

\setcounter{equation}{0}
Interaction with external time-dependent fields in semiconductor superlattices (SLs)
leads in many cases to completely new ways of electrical transport:
negative pumping of electrons, dynamical localization,
photoinduced electric field domains (EFDs),
and bistability between positive and negative current~[1,2].
In voltage biased weakly-coupled SLs, stationary EFDs arise if the well doping is large enough.
They usually consist of two homogenous electric field regions
separated by a domain wall (DW) of accumulated electrons.
When the carrier density decreases below a critical value,
the DW is unable to find a stable position
and starts to move over several periods. Its motion and recycling gives rise to
spontaneous self-oscillations of the electric current~[3].
As doping is experimentally hard to manipulate, this static-to-dynamic-state transition
has been achieved by the aplication of transverse magnetic fields~[4],
laser illumination in undoped SLs~[5], and careful variation of temperature~[6].

Due to the strongly nonlinear behaviour of a weakly-coupled SL
(stemming from the interplay between
tunneling processes and averaged Coulomb interactions),
perturbing the system with an ac potential involves
the rising of even more complex aspects in the physics of transport.
In particular, it has been reported analyses of time-dependent current
in the presence of a low-frequency signal both experimentally 
and theoretically [7]. In that case, the ac frequency is of the order
of tens of MHz and the ac potential results in an adiabatic modulation of the system.
In this paper we are interested in a very different regime.
We investigate the time-dependent current through a multiple quantum
well driven by a high-frequency ac potential, $V_i^{ac}(t)=V_i^{ac}\cos(\omega_0 t)$,
where the ac frequency $f_{ac}=\omega_0/2\pi$ is of the order of several THz
and $V_i^{ac}$ is the ac amplitude in the $i$th quantum well (QW).
It is well known that in this case photoassisted tunneling
takes place, and the electronic states develop
side-bands which act as new tunneling channels.

The Keldysh Green-function formalism [8] allows to obtain
general expressions for the tunneling current.
It also permits to include electron-electron
interaction in the system. Both the limits of noninteracting systems 
and local equilibrium can be deduced within this scheme.
We shall restrict ourselves to analyze the case of
a weakly-coupled SL where the electron-electron interaction will
be included in a mean-field (Hartree) manner.
From the time evolution of the occupation number operator
of the $i$th QW ($N_i$), the change in the number of electrons
in the $i$th QW is $e\langle\dot{N}_i\rangle=I_{i-1,i}-I_{i,i+1}$.
This continuity equation relates the change of the number of electrons in the $i$th QW
with the current density flowing from the $i-1$st ($i$th) QW
to the $i$th ($i+1$st) QW, $I_{i-1,i}$ ($I_{i,i+1}$).
By using the standard techniques of motion equation [8] one arrives to
\begin{eqnarray}
&& I_{i,i+1}(t)=\frac{2e}{\hbar}\mbox{Re}
\sum_{\vec{k}_{i} \vec{k}_{i+1}}T_{\vec{k}_{i} \vec{k}_{i+1}}
\int d\tau \left[ G^r_{\vec{k}_{i+1}}(t,\tau) g^<_{\vec{k}_{i}}(\tau,t)
+G^<_{\vec{k}_{i+1}}(t,\tau)g^a_{\vec{k}_{i}}(\tau,t) \right] \, ,\label{eq1}
\end{eqnarray}
$T_{\vec{k}_{i} \vec{k}_{i+1}}$ being the transmission coefficient.
$\vec{k}_{i}$ represents the set of quantum numbers which label the electronic states
in the $i$th QW. Here, $g^{a (<)}_{\vec{k}_{i}}$ is the advanced (lesser)
Green function (GF) for a decoupled QW in the presence of an ac potential and scattering processes.
$G^{r (<)}_{\vec{k}_{i}}$ is the retarded (lesser) QW GF when scattering,
ac signal and tunneling are present.
In the case of a SL, the GFs which appear in Eq.~(\ref{eq1}) can be
written down exactly by making several assumptions. 
Firstly, one considers the scattering
self-energy (due to, e.g., impurities or LO phonons) in a phenomenological
way as an energy indenpendent constant
(which is denoted here by $\gamma=\mbox{Im}\Sigma_{sc}$).
As in weakly-coupled SL's the scattering lifetime ($\sim 1$~ps)
is much shorter than the tunneling time ($\sim 1$~ns),
we can assume \emph{local equilibrium} within each QW
and neglect the tunneling self-energy in the expressions of the GFs.
The electron transport is thus \emph{sequential}.  
Once these approximations are made, the expression for the retarded (lesser)
ac-driven QW GF in the presence of scattering is
$G^{r(<)}_{\vec{k}_{i}}(t,t')=e^{ \frac{ieV_i^{ac}}{\hbar\omega_0}
\left(\sin\omega_0 t-\sin\omega_0 t'\right) } \bar{G}^{r(<)}_{\vec{k}_{i}}(t-t')$,
where $\bar{G}^{r(<)}_{\vec{k}_{i}}(t-t')$
is the static retarded (lesser) QW GF's in the presence of scattering.
Similar relations hold for $g^{a(<)}_{\vec{k}_{i}}(t,t')$.
By using these expressions the current in the case of local equilibrium reads~[9]
\begin{eqnarray}
&&J_{i, i+1}(t)=\frac{2e}{\hbar}\sum_{\vec{k}_{i} \vec{k}_{i+1}}
T_{\vec{k}_{i} \vec{k}_{i+1}}
\sum_{m=-\infty}^{m=\infty}J_m(\beta_{i,i+1}) \bigg\{
\cos\left(\beta_{i,i+1} \sin\omega_0 t-m\omega_0 t\right)
\\
\nonumber
&&\times\int d\epsilon\left[ A_{\vec{k}_{i+1}}(\epsilon+m\hbar\omega_0)A_{\vec{k}_{i}}(\epsilon)
\left(f_i(\epsilon)-f_{i+1}(\epsilon+m\hbar\omega_0)\right)\right]
+\sin(\beta_{i,i+1}\sin\left(\omega_0 t-m\omega_0 t\right)
\\
\nonumber
&&\times\int d\epsilon \left[ A_{\vec{k}_{i+1}}(\epsilon+m\hbar\omega_0)\mbox{Re}
\bar{g}^a_{\vec{k}_{i} \vec{k}_{i}}(\epsilon)f_{i+1}(\epsilon+m\hbar\omega_0)+
\mbox{Re} \bar{G}^r_{\vec{k}_{i+1} \vec{k}_{i+1}}(\epsilon+m\hbar\omega_0)
A_{\vec{k}_{i}}(\epsilon)f_i(\epsilon) \right] \bigg\} \, , \label{eq2}
\end{eqnarray}
where $A_{\vec{k}_{i}}$ denotes the spectral function
for the $i$th isolated QW in the presence of scattering.
$J_p$ is the $p$th Bessel function whose argument is given by
$\beta=e(V_{i}^{ac}-V_{i+1}^{ac})/\hbar\omega_0$
and $f_i(\epsilon)$ is the Fermi-Dirac distribution function for the $i$th QW.

\begin{figure}[ht]
\begin{minipage}[b]{0.5\textwidth}
\centering
\includegraphics[width=0.9\textwidth,clip]{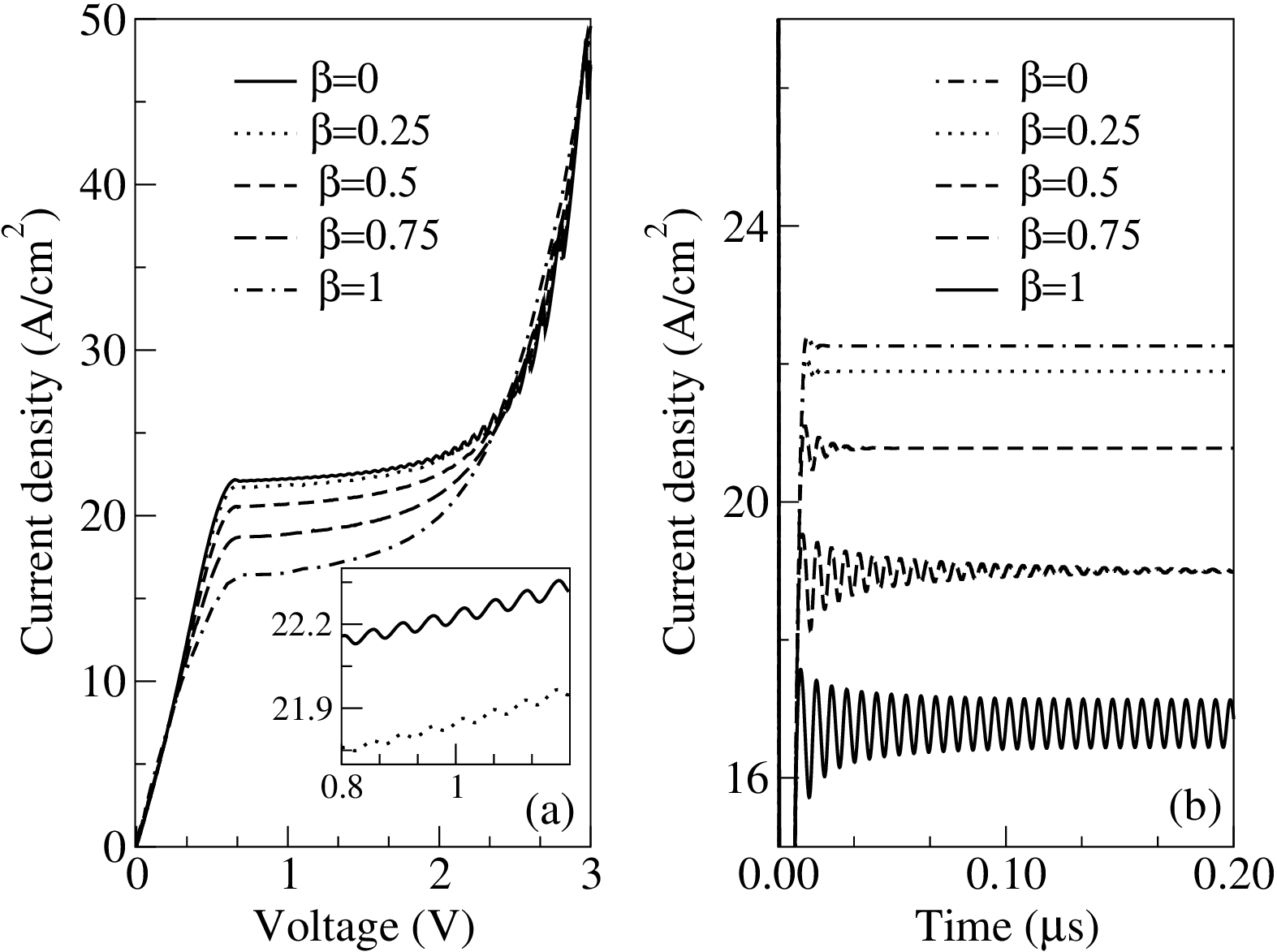}
\end{minipage}%
\begin{minipage}[b]{0.5\textwidth}
\centering
\includegraphics[width=0.9\textwidth,clip]{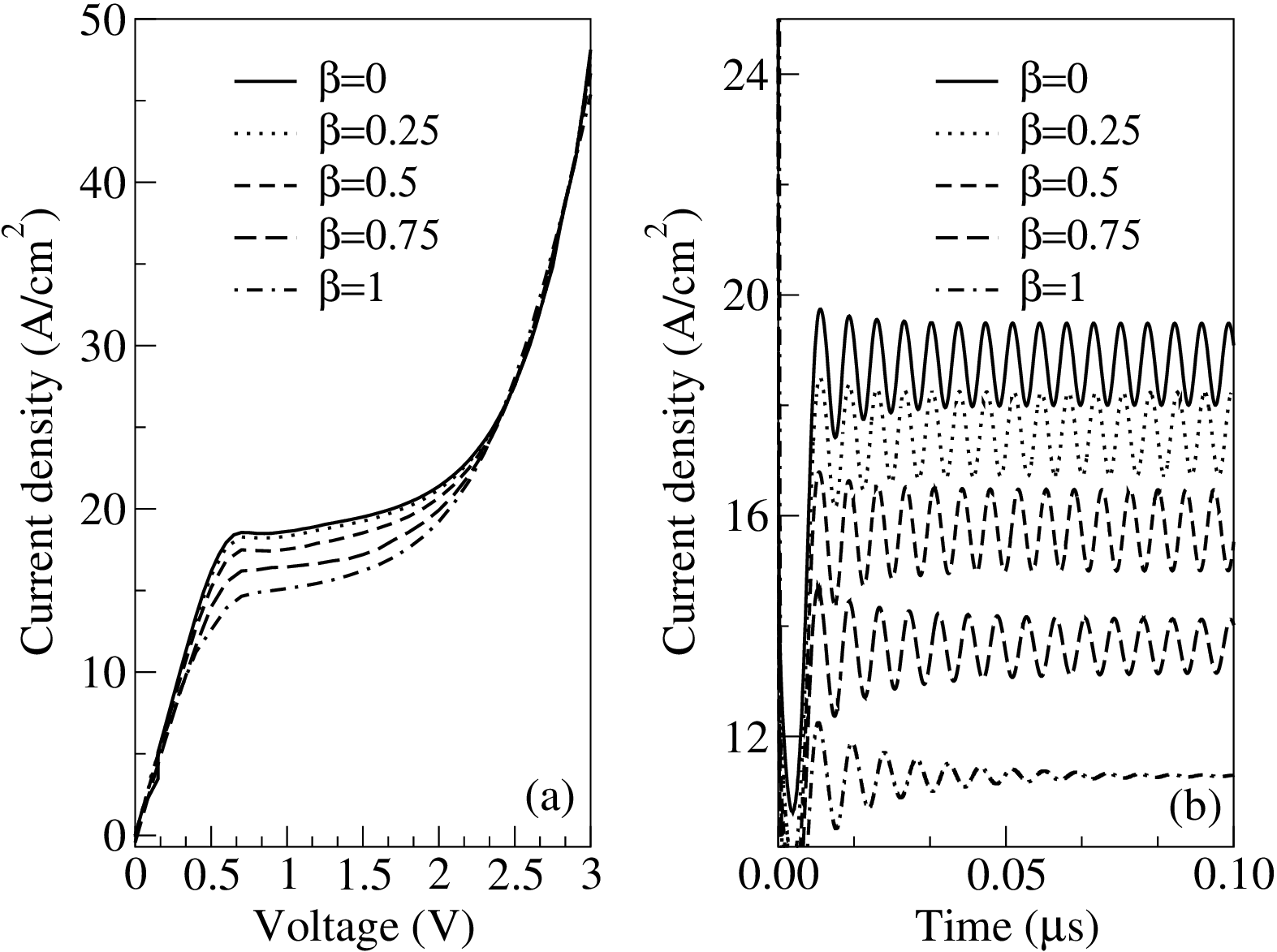}
\end{minipage}\\[-10pt]
\begin{minipage}[t]{0.5\textwidth}
\caption{(a) Time-averaged current vs. applied voltage (for parameters see text).
(b) Time dependence of the current for $V=1.1$~V
for the values of $\beta$ shown in (a).\label{fig1}}
\end{minipage}\hspace{0.4cm}%
\begin{minipage}[t]{0.47\textwidth}
\caption{(a) Time-averaged current vs. applied voltage as $\beta$ is varied.
(b) $I(t)$ for $V=1.1$~V. Curves for $\beta\neq0$ are shifted for clarity.\label{fig2}}
\end{minipage}%
\end{figure}

Notice that the current may be written as
$I(t)= I_0+\sum_{l>0} I^{\cos}_l \,\cos\left(l\omega_0\; t\right)
+I^{\sin}_l \,\sin\left(l\omega_0\; t\right)$,
where $I_0$ is the time-averaged current~[1].
$I^{\cos}_l$ and $I^{\sin}_l$ contain higher harmonics for $l>0$.
Now, since we are interested in the photoassisted current, $\omega_0$
is much larger than the tunneling rate.
In addition, the scattering lifetime
represents the lowest temporal cutoff above which
our assumption of local equilibrium within each QW holds.
In other words, to ensure the vality of Eq.~(\ref{eq2})
one is restricted to study dynamical
processes whose time variation is \emph{longer} than $\hbar/\gamma$.
As intense THz fields typically have $\hbar\omega_0>\gamma$
we must carry out a time average of $I(t)$. This implies that
the explicit time variation of $I(t)$ vanishes and
we are left with the implicit change of $I_0$ with respect to time.
This variation (in scales larger than $\hbar/\gamma$)
results from the evaluation of the continuity equation
for $i=1,\ldots,N$, where $N$ is the number of wells,
supplemented with Poisson equations, constitutive relations,
and appropiate boundary conditions (see [9] for details).
Of course, by incorporating an accurate microscopic model
for the scattering dynamics
and calculating the \emph{nonequilibrium} distribution function for each QW
in the presence of the ac signal, one could investigate
all the parameter range. This is out of the scope of the present work.

The total current traversing the sample consists of the tunneling plus displacement currents,
$I(t)=J_{i,i+1}+(\epsilon/d)(dV_i/dt)$,
where $(\epsilon$ is the static permittivity, $d$ the barrier width,
and $V_i$ the voltage drop in the $i$th barrier.
Given reliable initial conditions, $I(t)$ is self-consistently calculated
for a $N=40$ SL with 13.3-nm GaAs wells and 2.7-nm AlAs barriers.
Well doping is $2\times10^{-10}$~cm$^{-2}$ and we take $\gamma=8$~meV and $f_{ac}=3$~THz.
In Fig.~\ref{fig1}(a) the average of $I(t)$ is plotted as a function of the applied dc bias, $V$.
Without ac forcing the $I$--$V$ curve shows branches after the peak corresponding to resonant
tunneling between the lowest subbands (see inset).
This feature is distinctive of static EFD formation~[1-3].
In the presence of an ac signal the branches coalesce and a plateau is formed--
this is the key signature of current self-oscillations.
$I(t)$ for $V=1.1$~V is depicted in Fig.~\ref{fig1}(b).
For $\beta=0$ the current achieves a constant value after a transient time.
As $\beta$ is increased damped oscillations are observed until a stable
current self-oscillations arise when $\beta=1$. This is an indication that
the ac potential induces a transition from a stationary configuration towards a dynamic state
possibly via a supercritical Hopf bifurcation.
The oscillation frequency is of the order of 150~MHz, much smaller than $\omega_0$.
Then we conclude that the existence of sidebands
and its influence over the non-linear behaviour of the system
drives the SL towards oscillations.

Increasing $\gamma$ results in dramatic consequences. This may be effectively achieved by either
applying a transverse magnetic field~[4] or raising the temperature~[6].
In the absence of ac potentials for $\gamma=11$~meV
there is a voltage range ($\sim$0.8-1.3~V) where now self-oscillations show up~[3]
(see Fig.~\ref{fig2}(a)).
As $\beta$ enhances the plateau starts to be replaced by a positive differential resistance
region. There is a similar well-known phenomenon in weakly-coupled SLs: under a critical value of the
carrier density neither static nor moving DWs exist and the electric field drops
homogenously across the \emph{whole} sample.
In our case the doping density is constant and it is the ac field the physical
parameter which forces this transition.
To illustrate this we have calculated $I(t)$ for a fixed bias $V=1.1$~V (see Fig.~\ref{fig2}(b)).
At $\beta=0$ the current oscillates with a frequency of 170~MHz.
This is a result of the motion of the accumulating layer of electrons,
and its recycling in the highly-doped contacts~[3].
For $\beta=2$ the current is already damped,
and $I(t)$ reaches rapidly a uniform value for $\beta=4$.
This is a striking feature-- an oscilation disappearance induced by an ac potential.
A qualitative explanation follows.

The general transition (static EFDs)$\rightarrow$(moving EFDs)
$\rightarrow$(homogenous electric field profile)
depends basically on the particular shape of the drift velocity between two wells~[3-6].
For a sufficiently high peak-to-valley ratio the system evolves towards a static EFD situation.
If this ratio is very low the applied voltage drops smoothly across the sample.
In between, self-sustained current oscillations show up.
In our case the application of a THz field produces tunneling with absorption
and emission and photons, thus strongly modifying the current between two wells
(or, roughly, the drift velocity).
In particular, for the values used in the text the zero-photon tunneling peak
(weighted by $J_0^2(\beta)$) is depressed
and the current associated with the emission of, e.g., one photon becomes comparable
when $\beta$ increases from 0.
Then the peak-to-valley ratio
decreases as $\beta$ increases and the current oscillations are reached
provided we start from a stationary current (see Fig.~\ref{fig1}).
On the other hand, if the dynamic configuration is stable for $\beta=0$
the ac potential will tend to drive the SL to a
trivially homogenous electric field profile (see Fig.~\ref{fig2}).
A more detailed analysis will be given elsewhere~[9].

\mysection{Acknowledgments}
We thank R. Aguado for fruitful discussions. This work is supported by the Spanish DGES grant
PB96-0875 and by the European Union TMR contact FMRX-CT98-0180.

\mysection{References} 
\newlength{\mydescr}
\settowidth{\mydescr}{[1]}
\begin{list}{}%
{\setlength{\labelwidth}{\mydescr}%
\setlength{\leftmargin}{\parindent}%
\setlength{\topsep}{0pt}%
\setlength{\itemsep}{0pt}}
\item [\textrm{[1]}]
R. Aguado and G. Platero, Phys. Rev. Lett. {\bf 81}, 4971 (1998).
\item [\textrm{[2]}] 
B.J. Keay {\it et al.}, Phys. Rev. Lett. {\bf 75}, 4102 (1995).
\item [\textrm{[3]}]
J. Kastrup {\it et al.}, Phys. Rev. B {\bf 55}, 2476 (1997);
A. Wacker, M. Moscoso, M. Kindelan, and L.L. Bonilla, {\it ibid.} 2466 (1997).
\item [\textrm{[4]}]
B. Sun {\it et al.}, Phys. Rev. B {\bf 60}, 8866 (1999).
\item [\textrm{[5]}]
N. Ohtani, N. Egami, H.T. Grahn, and K.H. Ploog, Phys. Rev. B {\bf 61}, 5097(R) (2000).
\item [\textrm{[6]}]
J.N. Wang {\it et al.}, Appl. Phys. Lett. {\bf 75}, 2620 (1999);
D. S\'anchez, L.L. Bonilla, and G. Platero, cond-mat/0104117.
\item [\textrm{[7]}]
K.J. Luo {\it et al.}, Phys. Rev. Lett. {\bf 81}, 1290 (1998);
D. S\'anchez, G. Platero, and L.L. Bonilla, Phys. Rev. B {\bf 63}, 201306(R) (2001). 
\item [\textrm{[8]}]
H. Haug and A.-P. Jauho, \emph{Quantum Kinetics in Transport and Optics of Semiconductors}
(Springer-Verlag, Berlin, 1996).
\item [\textrm{[9]}]
R. L\'opez, D. S\'anchez, and G. Platero, in preparation (2001).
\end{list} 
\end{document}